\documentclass{article}

\usepackage{arxiv}

\usepackage[utf8]{inputenc} % allow utf-8 input
\usepackage[T1]{fontenc}    % use 8-bit T1 fonts
\usepackage{hyperref}       % hyperlinks
\usepackage{url}            % simple URL typesetting
\usepackage{booktabs}       % professional-quality tables
\usepackage{amsfonts}       % blackboard math symbols
\usepackage{nicefrac}       % compact symbols for 1/2, etc.
\usepackage{microtype}      % microtypography
\usepackage{lipsum}
\usepackage{graphicx}
\usepackage{amsmath}
\usepackage{amsthm}
\usepackage{lineno}
\usepackage{cite}

\graphicspath{ {./images/} }

\title{Mathematical Tri-State Model for Bee
Shimmering Propagation Dynamics}

\author{
 Navin Patel* \\
  School of Engineering and Materials Science\\
  Queen Mary, University of London\\ 
  Mile End Road, London E1 4NS\\
  United Kingdom \\
  \texttt{navin.patel@qmul.ac.uk} \\
   \And
Henri Huijberts \\
  School of Engineering and Materials Science\\
  Queen Mary, University of London\\ 
  Mile End Road, London E1 4NS\\
  United Kingdom \\
  \And
Kaspar Althoefer \\
  School of Engineering and Materials Science\\
  Queen Mary, University of London\\ 
  Mile End Road, London E1 4NS\\
  United Kingdom \\
    \And
Ketao Zhang* \\
  School of Engineering and Materials Science\\
  Queen Mary, University of London\\ 
  Mile End Road, London E1 4NS\\
  United Kingdom \\
  \texttt{ketao.zhang@qmul.ac.uk} \\
}

\begin{document}
\maketitle
\begin{abstract}
Bees undergo a self-organised process known as shimmering, where they form emergent patterns when they interact with each other on the nest surface as a defence mechanism in response to predator attacks. Many experimental studies have empirically investigated how the transfer of information to neighbouring bees propagates in various shimmering processes by measuring shimmering wave strength. However, there is no analytical modelling of the collective defence mechanism in nature. Here we introduce the first analytical tri-state Inactive-Active-Relapse (IAR) model to formulate the intrinsic process of bee shimmering. The major shimmering behaviour is shown to emerge under theoretical conditions which is demonstrated numerically and visually by simulating 1,000,000 bee agents, while the number of agents is scalable. Furthermore, we elaborate on these mathematical results to construct a wave strength function to demonstrate the accuracy of shimmering dynamics. The constructed wave strength function can be adapted to peak between 50-150ms which supports the experimental studies.  Our results provide a foundation for further theoretical understanding of bee shimmering wave dynamics and could serve as inspiration for modelling other self-organised phenomena across scientific applications.
\end{abstract}

% keywords can be removed

\section{Introduction}
Shimmering in giant honeybees (Apis dorsata) \cite{speedingsocialwaves} is a defensive mechanism whereby bees at the nest surface respond with an anti-predatory impact against hornets \cite{speedingsocialwaves,decisionMaking,StereoMotion}. This collective behaviour involves each individual bee expanding its abdomens outwards in a continuous and linear response~\cite{BeeVibrate,BeeBook,BeeDefStrat,Assam,BeeSelfAssem,BeeKinematics}. These complex social waves form a self-organised system~\cite{BeeSelfAssem,BeeStochSyc} similar to mexican waves (stadium waves)~\cite{speedingsocialwaves,decisionMaking,StereoMotion}, which propagates across the nest surface~\cite{BeeDefStrat} in a fraction of a second.
Ecologists have studied the propagation mechanism in shimmering by observing the time differences from various recorded videos, and heat maps and observing the transfer of information from an individual honeybee to the others~\cite{speedingsocialwaves,decisionMaking,StereoMotion,BeeVibrate,BeeBook,BeeDefStrat,Assam,BeeSelfAssem,BeeKinematics}. Image acquisition and segmentation techniques have been used to observe the decision-making processes of honeybees within shimmering~\cite{decisionMaking,StereoMotion,SocialRepel}. Computer vision techniques such as stereoscopic motion analysis have also been used to observe the positional differences of the bees during shimmering ~\cite{StereoMotion}. 

Further studies have also been done to observe this self-organisation process by assessing the phase transition from individual bee abdomen flickering behaviour to synchronous flickering otherwise known as shimmering~\cite{BeeStochSyc}. Other studies are based on nest vibrations during shimmering~\cite{BeeVibrate}. The findings of the above work provide a strong foundation for the biological understanding of this self-organised process by empirically measuring various shimmering behaviours. Although these studies provide solid observations of the effects of shimmering, there still lacks a unifying theory that supports any experimental studies to understand the underlying dynamics of shimmering. By developing theoretical foundations, one can go beyond the restrictions of experimental work. For example, the scalability of the number of bees can be freely varied and the desired effects of different shimmering waves can be obtained without relying on physical bees and wasps. 
 
 \begin{figure}[ht!]
\begin{center}
    \includegraphics[width=1\textwidth]{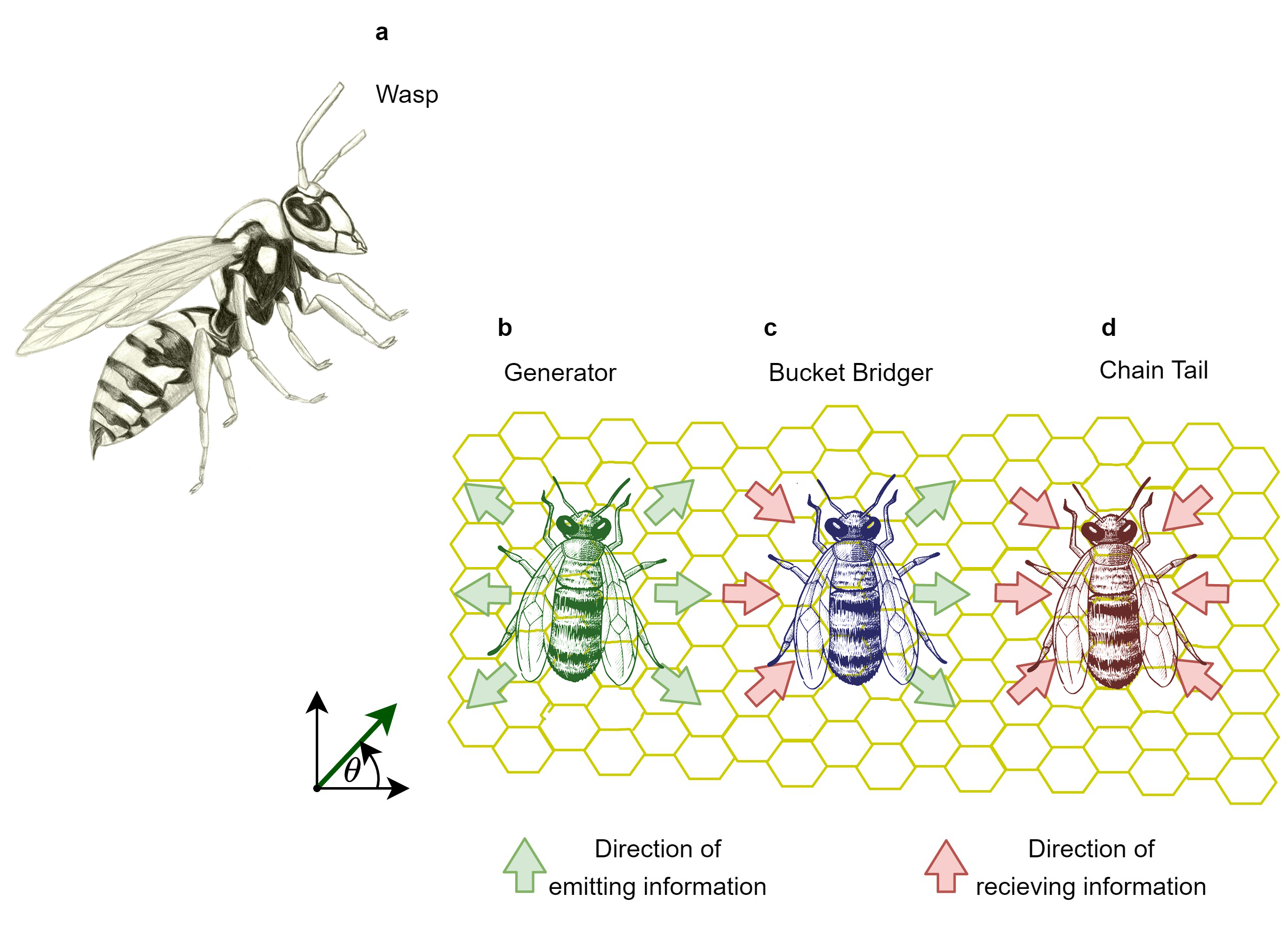}
\end{center}
  \linespread{1}\selectfont
\caption{{\bf Bee shimmering agents information transfer}
Bee shimmering takes place on the surface of the bee nest and is initiated when a wasp (a) is in close vicinity. There are three possible types of agents during shimmering known as generator (b), bucket bridger agents (c), and chain tail (d). The red arrows show the origin of the transfer of information, the green arrows determine where the information is being propagated to. The bucket bridger agents (c) acts as both an emitter and a receiver of information. The chain tail agents (d) do not respond to information in any way, whereas the generator agents (b) act as the transmitters to start the wave. Each agent has its own respective neighbourhood angle of where the wave would be received as indicated in the reference axis. The green arrow indicates the direction from which the wave is originating, pointing towards the nearest bee transmitting the signal.
}
\label{fig1}
\end{figure} 

There are three possible types of agents during the bee shimmering process\cite{speedingsocialwaves,decisionMaking,StereoMotion} as shown in Fig.~\ref{fig1}. These are known as the bucket bridger, chain tail and generator agents respectively\cite{speedingsocialwaves}. These three agents alter the dynamics of shimmering depending on the proportion of these individual agents affecting the transfer of information. The signal for which bees interact is a combination of visual cues and through a process known as nasonov pheromone~\cite{speedingsocialwaves,decisionMaking,StereoMotion}, where a scent is released by bees to send this signal. Nasonov pheromone is a local process, meaning bees that are exposed to the scent interact with their neighbours, whereas visual cues can trigger long-range interactions with the possibility of triggering bees in neighbouring nests. In shimmering, there are major shimmering (dubbed as the shimmering wave), and smaller wavelets known as saltatoric process and bucket bridging (minor shimmering)~\cite{speedingsocialwaves,decisionMaking,StereoMotion}. These waves depend on the characteristics of the bees that are around the area of occurrence, the stages of luring a wasp and the reach of nasonov pheromone~\cite{Nasonov}. On a nest surface, it can be shown that a single bee is surrounded by six to ten others on average~\cite{speedingsocialwaves,BeeDefStrat,Assam,BeeStochSyc}. However, upon hive edges and multiple bee layers, the average number is uncertain and the average neighbouring quantity of the bees is left as a varying parameter in this study due to the lack of evidence. 

Prior to this work, there have been no mathematical models which describe the propagation dynamics of bee shimmering. Here we propose an analytical model which categorises the behaviour of bees into various states and describes how each of these states changes over time. The different types of interactions that can occur are considered as well as a relationship of how these interactions affect the strength and type of waves. Since experimental studies do not provide enough underlying theory to transfer ideas from one scientific field to another in a multidisciplinary manner. The proposed theoretical model provides scientists and engineers with a starting point to create bee shimmering inspired applications.

\section{Bee Shimmering Propagation Dynamics}

\subsection{State interactions}

Based on extensive empirical studies of bee shimmering~\cite{speedingsocialwaves,decisionMaking,StereoMotion,BeeVibrate,SocialRepel,BeeBook,BeeDefStrat,Assam,BeeSelfAssem,BeeKinematics}, the inspiration of the Mexican wave modelling~\cite{MexicanWaves,MexicanWaves2,MexicanWaves3}, and other self-organised nature systems~\cite{AnimalInter,Starling1,SelfOrgBook,SyncBook,ComplexityBook,Starling2,Starling3,Starling4,Starling5,RecNN}, it evident that there are three possible types of agents (Generator, Bucket Bridger, and Chain Tail)~\cite{speedingsocialwaves,decisionMaking,StereoMotion} as illustrated in Fig.~\ref{fig1} and a single bee can be in only three possible states (Active, Inactive and the Relapse states).

\begin{figure}[!ht]
\begin{center}
    \includegraphics[width=0.81\textwidth]{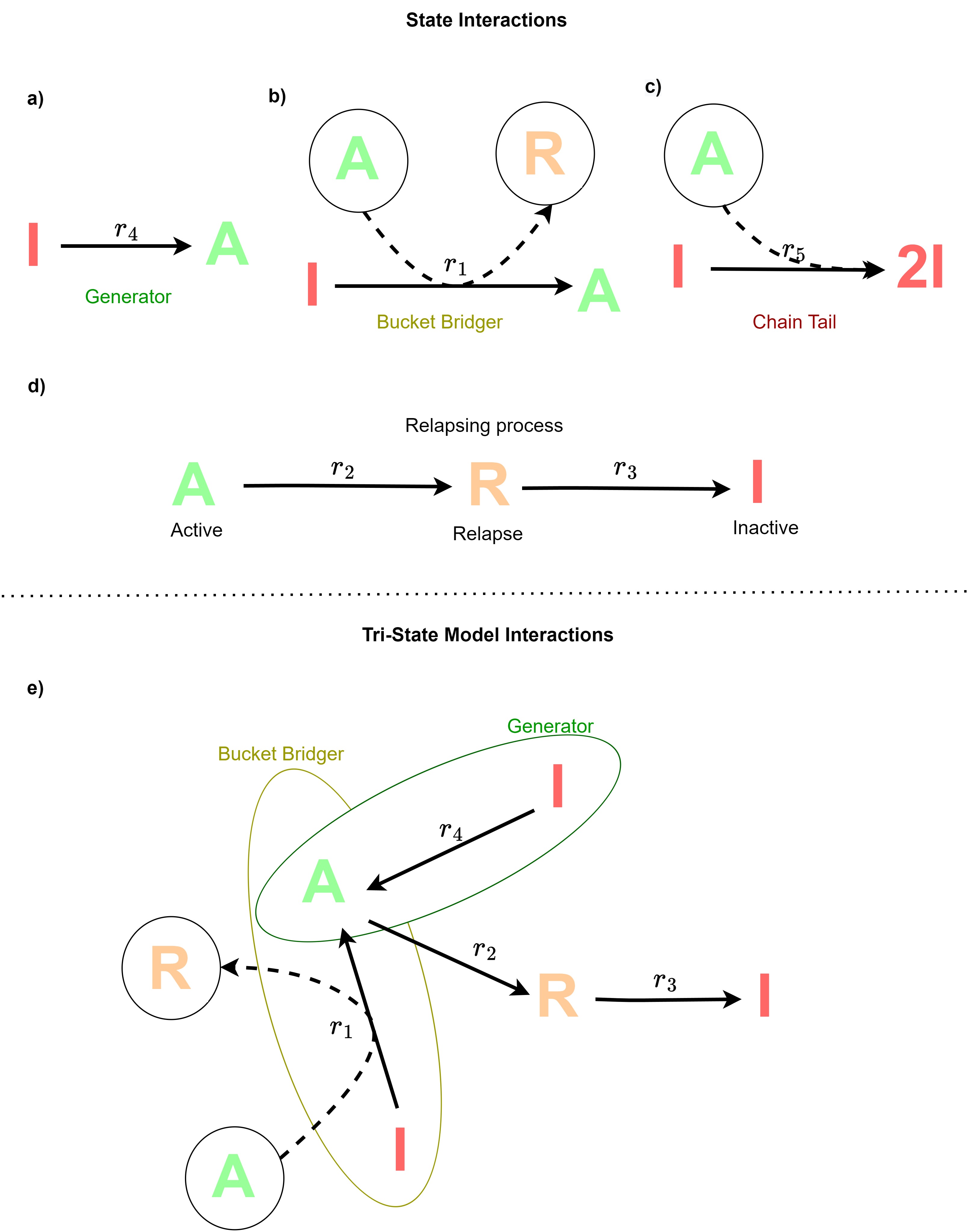}
\end{center}
  \linespread{1}\selectfont
\caption{{\bf State transitions of the various agents}
Three states have been classified, inactive, active and relapse states. These states classify the motion of each of the bees. Inactive states indicate the state at which bees are not expanding their abdomens, the active state classifies the process at which the bee goes from an inactive state to its maximum expansion. Thirdly the relapse state is the process during which the bee returns to the inactive state from its maximum flexion. Reaction (d) indicates the stages at which active state Bees return to the inactive state. The reactions showcase the interaction dynamics of each individual type of agent, which corresponds to generator agent (a), bucket bridger agents (b), and chain tail agents (c), and the relapsing process (d). Diagrams (a)-(d) show the state transitions for the state interaction model. Diagram (e) shows the complete tri-state model interactions showcasing the full interaction dynamics of bee shimmering showing a redundancy of the chain tail agent where the rate $r_{5}$ is encapsulated when the rates $r_{1}$ from (b) and $r_{4}$ (a) has a probability for the reactions not being undergone.}
\label{fig2}
\end{figure}

The inactive state ($\mathcal{I}$ state) refers to agents that are at rest, meaning their abdomens are not stretched. These agents can be influenced by the active agents during interactions~\cite{speedingsocialwaves,decisionMaking,StereoMotion,BeeVibrate,SocialRepel,BeeBook,BeeDefStrat,Assam,BeeSelfAssem,BeeKinematics}. The active state ($\mathcal{A}$ state) indicates an agent is either stretching its abdomen to its maximum capacity or has already done so. The agents in this state play a role in determining the strength of the wave and influencing neighbouring inactive agents. The relapse state ($\mathcal{R}$ state), although not explicitly defined in previous works, is considered in this model to account for the time it takes for an active agent to return to an inactive state. Both active and relapse states assume that these agents are not influenced by neighbouring agents. However, unlike active agents, relapse agents do not influence other agents. 

% Agents can be grouped into three categories: generator, bucket bridger, and chain tail agents, as illustrated in Fig.\ref{fig1}. 
Fig.\ref{fig2} depicts the evolution of each individual agent using straight arrows, while the effects of interactions with external agents are represented by dashed arrows. The states of the external agents are indicated by circles. If the external agent is in the active state, the interactions shown in Fig.\ref{fig2} will take place, regardless of its category. The relapsing process in Fig.\ref{fig2}(d) illustrates what happens when active agents return to the inactive state. All state changes are governed by reaction rates $r_{1}$, $r_{2}$, $r_{3}$, $r_{4}$, and $r_{5}$, similar to the mass action law~\cite{MassAction}. Each agent in the model is defined through a short period of time of up to $1000ms$, they can only have a single type (generator, bucket bridger, and chain tail) within this period. Given a longer period, they can change their types in the physical system. Each agent can only be in any single state (inactive, active, relapse) at any given point in time. The agent states and categories are the individual microscopic properties of the bee shimmering phenomena. However, within the overall macroscopic scale of the shimmering, there are two phases to shimmering, these are known as the pre-stroke and post-stroke phases.

The pre-stroke phase is a strategy employed by bees in which they intentionally instigate a wasp into a chase ~\cite{speedingsocialwaves,decisionMaking,StereoMotion}. This provocation leads the wasp to pursue the bees, luring the wasp to the bee nest. Subsequent to the chase, the post-stroke phase initiates, marking the onset of major shimmering. Throughout this shimmering process, individual bees release a unique scent known as the Nasonov pheromone when they extend their abdomens or elevate their wings, a state referred to as active. Bees in the immediate vicinity of this scent, initially in an inactive state, become active and reciprocate this behaviour ~\cite{Nasonov}. This leads to a diffusion of the scent across the nest, thereby causing the shimmering to manifest in two distinct ways: minor and major shimmering ~\cite{speedingsocialwaves,decisionMaking,StereoMotion}.

Major shimmering consists of the large waves produced, whereas minor shimmering includes small wave formations and is further subdivided into the saltatoric process and the bucket bridging process ~\cite{speedingsocialwaves,decisionMaking,StereoMotion}. Notably, the saltatoric process, despite its relatively lower wave strength, rapidly produces multiple small wavelets, thereby contrasting with the major shimmering wave and the slower bucket bridging process. On the other hand, the bucket bridging process, while slower in wave speed, generates larger clusters and fewer wavelets compared to the saltatoric process. Nevertheless, the clusters formed during the bucket bridging process do not attain sizes observed in the major shimmering phenomena. These dual manifestations of minor shimmering underscore the intricate and multifaceted nature of the bees' defensive behaviour ~\cite{speedingsocialwaves,decisionMaking,StereoMotion,BeeVibrate,SocialRepel,BeeBook,BeeDefStrat,Assam,BeeSelfAssem,BeeKinematics}. Having established the relevant biological processes during what occurs on both the microscopic and macroscopic scales, these factors are taken into consideration for the modelling assumptions. By fusing these biological findings with the notion of state interactions, we devise a mathematical model for shimmering waves. We can then predict and simulate the emergence of major and minor shimmering waves, providing detailed insights into the emergent properties as dictated by the numbers of active, inactive, and relapse agents.

\subsection{State interaction assumptions}

The diagrams provided in Fig.~\ref{fig2} follow a multitude of assumptions based on biological and ecological studies of bee shimmering ~\cite{speedingsocialwaves,decisionMaking,StereoMotion,Assam,SocialRepel,BeeVibrate}.

\begin{enumerate}

\item This study focuses only on the post-stroke phase of interactions and does not consider the pre-stroke phase of shimmering~\cite{speedingsocialwaves,decisionMaking,StereoMotion}.
\item The bee shimmering communication structure is mainly caused via a Nasonov pheromone scent ~\cite{Nasonov} as well as a combination of localised visual cues and nest vibrations~\cite{speedingsocialwaves,decisionMaking,StereoMotion,Assam,SocialRepel,BeeVibrate}. The interactions considered in this study are limited to those between neighbouring agents and do not include those between generator agents and the major shimmering wave from a distance. 
\item The number of total agents remains constant throughout the study. This assumption is made despite evidence that the number of chain tail, generator, and bucket bridger agents can change~\cite{speedingsocialwaves,decisionMaking,StereoMotion,Assam,SocialRepel}. 
\item This study only considers the first major shimmering wave, as opposed to subsequent waves that may occur after the initial shimmer. The timescale of the first wave is limited to a maximum of $1000ms$~\cite{speedingsocialwaves,decisionMaking,StereoMotion} to minimize the risk of bees moving away from the hive, which would compromise the assumption that the number of agents stays constant.
\item  The saltatoric process and bucket bridging processes will be considered as minor shimmering initially. Major shimmering waves will be considered as shimmering waves. 

\end{enumerate}

Based on these assumptions, the diagrams in Fig.~\ref{fig2}(a)-Fig.~\ref{fig2}(d) can be transformed into equivalent chemical reactions of interactions~\cite{MassAction}.

\subsection{State interaction model}

Here we model the interactions in Fig.~\ref{fig2}(a)-Fig.~\ref{fig2}(d) by converting them into equivalent chemical reactions \cite{MassAction,ChemMod1,ChemMod2,ChemMod3,MathBio} using the law of mass action as shown below.

\begin{equation}\label{q}
\begin{array}{c}

I \stackrel{r_{4}}{\rightarrow} A\\

A+I \stackrel{r_{1}}{\rightarrow} R+A\\

I+A \stackrel{r_{5}}{\rightarrow} 2 I \\

A \stackrel{r_{2}}{\rightarrow} R \stackrel{r_{3}}{\rightarrow} I\\
\end{array}
\end{equation}

From the above chemical equations Eq.(\ref{q}), a set of first-order differential equations can be derived describing how the number of agents in the respective $\mathcal{I}$, $\mathcal{A}$ and $\mathcal{R}$ states change over time by defining a state vector $x$, a stoichiometry matrix $\Gamma$ with a chemical speed vector $\omega$, such that $\dot{x} = \Gamma \omega$ \cite{MassAction,ChemMod1,MathBio}.

\begin{equation}\label{t}
\dot{x}=\left[\begin{array}{c}
\dot{I} \\
\dot{A} \\
\dot{R}
\end{array}\right]=\left[\begin{array}{ccccc}
-1 & 0 & 1 & -1 & 1 \\
0 & -1 & 0 & 1 & -1 \\
1 & 1 & -1 & 0 & 0
\end{array}\right] \cdot\left[\begin{array}{c}
A I r_{1} \\
A r_{2} \\
R r_{3} \\
I r_{4} \\
A I r_{5}
\end{array}\right]
\end{equation}

Performing the corresponding matrix multiplication derives the following first-order differential equations: 

% \begin{equation}\label{10}
% \begin{array}{c}
% \dot{I}=-A I r_{1}+R r_{3}-I r_{4}+A I r_{5}\\

% \dot{A}=-A r_{2}+I r_{4}-A I r_{5}\\

% \dot{R}=A I r_{1}+A r_{2}-R r_{3}
% \end{array}
% \end{equation}

% Using the law of mass action \cite{MassAction,ChemMod1,ChemMod2,ChemMod3,MathBio}, the state interaction model is derived (as detailed in \textbf{Supplementary Information}) and governed by the following first-order differential equations: 

\begin{equation}\label{1}
\begin{array}{c}
\dot{I}=-A I r_{1}+R r_{3}-I r_{4}+A I r_{5}\\

\dot{A}=-A r_{2}+I r_{4}-A I r_{5}\\

\dot{R}=A I r_{1}+A r_{2}-R r_{3}
\end{array}
\end{equation}

Here we let $ S = \{\mathcal{I}, \mathcal{A}, \mathcal{R}\}$, denote the set of states, where 
$\mathcal{I}$, $\mathcal{A}$, and $\mathcal{R}$ correspond to the 'Inactive', 'Active', and 'Relapse' states, respectively. The parameters $I$, $A$ and $R$ are the total number of bee agents that are in the Inactive ($\mathcal{I}$), Active ($\mathcal{A}$) and Relapse ($\mathcal{R}$) states respectively. Where $r_1, r_2, r_3, r_4, r_5 \in \mathbb{R}_{\geq 0}$ and $\mathbb{R}_{\geq 0}=\{x \in \mathbb{R}: x \geq 0\}$. Using the diagrams in Fig.~\ref{fig2}(a)-Fig.~\ref{fig2}(d), by identifying the rates that induce ($r_{1}$ and $r_{4}$)and remove ($r_{2}$ and $r_{5}$)the active state, we define the basic shimmering threshold as $R_{0} = \frac{r_{1}+r_{4}}{r_{2}+r_{5}}$. When there are no chain tail agents in a system, this can be reduced to $R_{0} =\frac{r_{1}+r_{4}}{r_{2}}$. The numerator is the sum rates of inducing the active states, whereas the denominator is the sum rates of agents leaving the active states. The state interaction model Eq.~(\ref{1}) is simulated to visualise major and minor shimmering on a grid via neighbourhood interactions without the need of spatial coordinates. Each agent changes its state according to the interactions shown in Eq.~(\ref{q}) which are iterated through time steps.

\subsection{The tri-state IAR model}

In Fig.~\ref{fig1} there is a depiction of how information is transferred amongst the three types of bee agents. To take into account the types of agents, we first define a network of bees $\mathcal{N}(N,L)$ consisting of $N$ total bee agents and $L$ links formed via bee interactions. Then we define new variables describing the average number of bucket bridger and generator neighbours a selected agent has. This is known as the average degree. The average degrees of the bucket bridger and generator agents would be introduced as, $\left\langle k_{B}\right\rangle$ and $\left\langle k_{G}\right\rangle$ respectively. A single bee agent can be found with six to ten other bee agents on average~\cite{speedingsocialwaves,BeeDefStrat,Assam,BeeStochSyc}. Physically, for example, if we have $\left\langle k_{B}\right\rangle=6$ and $\left\langle k_{G}\right\rangle=1$ this would mean that a randomly selected bee agent would have six bucket bridger and one generator agents on average. Therefore, a feasible range for $\left\langle k_{B}\right\rangle$ would be between six and ten, and for $\left\langle k_{G}\right\rangle$ would be between zero and one ~\cite{speedingsocialwaves,decisionMaking,StereoMotion}. In turn $\left\langle k_{C}\right\rangle$ would be the average degree of the chain tail agents. However, from the definitions of what chain tail agents are capable of~\cite{speedingsocialwaves,decisionMaking,StereoMotion} and illustrated in both Fig.~\ref{fig1} and Fig.~\ref{fig2}(e), these agents are redundant. In other words, they do not do anything to pass the propagation of information, they are termed as impaired agents, as they prevent information from travelling. Thus the variables $\left\langle k_{B}\right\rangle$ and $\left\langle k_{G}\right\rangle$ should be introduced, whereas the chain tail agents could be modelled as a percolation~\cite{Percolation1,Percolation2,Percolation3}, of damaged nodes in the network as shown in Fig.~\ref{fig2}(e). Therefore the network of agents within the bee shimmering time frame only consists of bucket bridgers and generator agents, where the failed bucket bridger and generator agents are chain tail agents. When $r_{1}$ and $r_{4}$ fail, these agents are not induced into the active state and ultimately stay in the inactive state.

Fig.~\ref{fig2} (d) shows the proposed dynamics of the system. It consists of bucket bridger and generator agents and describes the process of returning the overall system to the $\mathcal{I}$ state. To consider the chain tail agents in the diagram, we assume that $r_{1}$ and $r_{4}$ would have respected pass or fail probabilities. If either $r_{1}$ or $r_{4}$ have failed then these agents will be classed as chain tail agents. To define this variable, we introduce the concept of transmissibility which is similar to SIR-type models \cite{Epidem1,SIRstab,MathBio,ComplexityBook}. The transmissibility $\lambda$ is the probability of inducing the active state of an inactive agent (whether from an inactive generator or bucket bridger agents) through an inactive and active ($I+A$) interaction of bucket bridgers. Therefore, $1-\lambda$ is the probability of remaining in an inactive state after an $I+A$ interaction caused by an active generator agent or active bucket bridgers. The total population of bees is defined as $N$. To incorporate stochastic processes into the model, the variables $I$ $A$ $R$ will be denoted by, $i$ $a$ and $r$ such that:

\begin{equation}\label{11}
  i=\lim _{N \rightarrow \infty} \frac{I}{N}\\
\end{equation}

\begin{equation}\label{12}
\quad a=\lim _{N \rightarrow \infty} \frac{A}{N}\\
\end{equation}

\begin{equation}\label{13}
\quad r=\lim _{N \rightarrow \infty} \frac{R}{N}  
\end{equation}

The indication of the average proportion of active agents amongst bucket bridgers is $\alpha\left\langle k_{B}\right\rangle a(t)$ where $\alpha$ is a constant between 0 and 1 \cite{speedingsocialwaves}. Therefore the inactive probability after an active bucket bridger contacts all the inactive bees around it is $(1-\lambda)^{\alpha\left\langle k_{B}\right\rangle a(t)}$. As a result, the derivation of the probability of an active agent occurring after active bucket bridger agents contact inactive bees around it will be given by $r_{1}$ which follows as: 

\begin{equation}\label{14}
r_{1}(t)=1-(1-\lambda)^{\alpha\left\langle k_{B}\right\rangle a(t)}
\end{equation}

Another proposed assumption here is a similar derivation for $r_{4}$ describing the rates of generator agents becoming active where the transmissibility $\lambda$ is the same for bucket bridgers and generator agents. Each active agent regardless of type has the same probability of conversing with active agents. Hence a similar process can be used to derive $r_{4}$ which follows as:  

\begin{equation}\label{15}
r_{4}(t)=1-(1-\lambda)^{\beta\left\langle k_{G}\right\rangle a(t)}
\end{equation}

Here $\beta$ is also a factor independent of $\alpha$ between 0 and 1. A similar derivation can be found for the rate $r_{1}$. Replacing $I$, $A$, $R$ for the variables $i$, $a$, $r$ and by substituting the equations for $r_{1}$ and $r_{4}$ into the state interaction model Eq.(\ref{1}), we obtain the tri-state IAR model. The proposed bee shimmering tri-state IAR model is shown as follows:

\begin{equation}\label{2}
\frac{d i}{d t}=-a i\left[1-(1-\lambda)^{\alpha\left\langle k_{B}\right\rangle a(t)}\right]+(1-i-a) r_{3}-i\left[1-(1-\lambda)^{\beta\left\langle k_{G}\right\rangle a(t)}\right]
\end{equation}

\begin{equation}\label{3}
\frac{d a}{d t}=-a r_{2}+i\left[1-(1-\lambda)^{\beta\left\langle k_{G}\right\rangle a(t)}\right]
\end{equation}.

% The derivation of the proposed IAR model is found in the \hyperlink{IARmodel}{IAR derivation in the \textbf{Appendix} section.} 
The proposed model is established assuming that the agent size $N$ is constant where no bees fly or move away from the hives as $i(t)+a(t)+r(t)=1$. The rates $r_1, r_2, r_3, r_4$ are now kept as probabilities such that $r_1, r_2, r_3, r_4 \in[0,1]$. The rate $r_5$ has been omitted as the rates $r_1$ and $r_4$ encapsulate the \hyperlink{IARmodel}{definition of the proportions of chain tail agent.} The variables $\alpha, \beta \in[0,1]$ are normalisation constants, and $\lambda$ is the probability of \hyperlink{IARmodel}{conversing an inactive agent to the active state.}

\subsection{Invariance and stability}

Having obtained this model, it is now of interest to determine its steady states, invariance properties and stability conditions and physical implications.

By renaming ${x}_{1}=i$, ${x}_{2}=a$, and by letting $\gamma=\alpha\left\langle k_{B}\right\rangle\ln(1-\lambda)^{-1}$ and $\delta=\beta\left\langle k_{G}\right\rangle\ln(1-\lambda)^{-1}$, Eq.(\ref{2})-Eq.(\ref{3}) can be re-written as:

\begin{equation}\label{4}
\left(
\begin{array}{c}
  \dot{x}_1 \\ \dot{x}_2
\end{array}
\right)=
\left(
\begin{array}{c}
   -x_{1} x_{2}\left(1-e^{- \gamma x_{2}}\right)+r_{3}\left(1-x_{1}-x_{2}\right)- 
x_{1}\left(1-e^{-\delta x_{2}}\right) \\
-r_{2} x_{2}+x_{1}\left(1-e^{-\delta x_{2}}\right)
\end{array}
\right)=:f(x_1,x_2)
\end{equation}

Let $\mathcal{S}=[0,1]\times [0,1]$ be the unit square. Note that it follows from the previous subsection that we need $0\le x_1\le 1,~0\le x_2\le 1$, which means that $\mathcal{S}$ is a positively invariant set~\cite{DynamicalSystems1} for Eq. (\ref{4}). The following proposition provides conditions for invariance.

\textbf{Proposition 1.} \emph{$\mathcal{S}$ is a positively invariant set for Eq. (\ref{4}) if and only if $r_2\ge 1-\mathrm{e}^{-\delta}$.} (\textbf{Appendix} \hyperlink{proof1}{Proof for Proposition 1.})

%Note that the reader is strongly encouraged to read the proof for all three propositions as shown in the \hyperlink{proof1}{methods section}. 

\hypertarget{Prp2}{\textbf{Proposition 2.}}  \emph{When $r_2\ge \delta$, the minor shimmering point $(1,0)$ is a globally asymptotically stable steady state for (\ref{4}), i.e., every solution of (\ref{4}) starting on $\mathcal{S}$ converges to $(1,0)$ for $t\to+\infty$.} (\textbf{Appendix} \hyperlink{proof2}{Proof for Proposition 2.})

 %\hyperlink{proof2}{in the methods section}.

\hypertarget{Prp3}{\textbf{Proposition 3.}} \emph{For all $1-\mathrm{e}^{-\delta}\le r_2<\delta$, there exists a unique globally asymptotically stable steady state of (\ref{4}) on ${\mathcal{S}}^*=[0,1]\times(0,1]$.} (\textbf{Appendix} \hyperlink{proof3}{Proof for Proposition 3.})

%Proof for proposition 3 can be found \hyperlink{proof3}{below proof 2} and is encouraged to read before proceeding further.

All propositions imply that the number of active and inactive agents converges to an equilibrium point depending on the parameters $r_{2}$ and $\delta$. For the equilibrium in $\mathcal{S}$, a small proportion of agents will be active until the next major shimmering wave occurs within 1 second. In addition, the stability conditions are dependent on $\delta$ defined in terms of $\langle k_{G} \rangle$ in rather than $\gamma$ in terms of $\langle k_{B} \rangle$ as bucket bridger agents do not start waves of any type. They only continue the propagation of information where generator agents start the wave.

In addition to the tri-state IAR model, the wave strength and further classification of the different types of shimmering waves can be deduced by taking into consideration the rates at which the active agents get conversed. 

\subsection{Wave strength function}
To demonstrate the flexibility of the tri-state IAR model Eq.(\ref{2})-Eq.(\ref{3}), a derivation of the wave strength is made and validated by comparing it to other work~\cite{speedingsocialwaves,decisionMaking,StereoMotion,SocialRepel,BeeVibrate} where wave strength has been measured empirically through experimental studies. The wave strength of the system is defined in accordance with the literature \cite{speedingsocialwaves,decisionMaking,StereoMotion,SocialRepel,BeeVibrate}. Throughout the derivation, the wave strength in this model is defined as the total rate at which the active states of the agents are induced. This definition is consistent with the comparisons as empirical studies suggest that the wave strength is defined by the characteristic of luminance changes across camera pixels that detect heat as the bees induce shimmering motion via image analysis software~\cite{speedingsocialwaves,decisionMaking,StereoMotion}. The wave strength definition in this modelling approach considers the rate at which agents change from inactive to active states, which is equivalent to luminance change. Therefore, the wave strength, $M(t)$, is defined as the relative wave strength among the population of agents over time. $M(t)$ is given by the summation of $r_{1}$ and $r_{4}$ as follows: 

\begin{equation}\label{5}
M(t)=r_{1}(t)+r_{4}(t)=2-\left[(1-\lambda)^{\alpha\left\langle k_{B}\right\rangle a(t)}+(1-\lambda)^{\beta\left\langle k_{G}\right\rangle a(t)}\right]
\end{equation}

 The validation of this function can be seen in the results below. Varying the parameters can produce results for both major and minor shimmering. An extension to this function would be to relate the transmissibility $\lambda$ to the directional control in shimmering. 
 
 \subsection{Wave strength function with directional control}

The directional propagation of the wave describes the transfer of information, dependent on the participation of agents (the number of active agents), from one particular direction to the other. Following the directed triggered hypothesis~\cite{speedingsocialwaves}, it is given that:

\begin{equation}\label{6}
P(\theta)=N_{r e l} \times \sin ^{2}\left(\theta\right)
\end{equation}

 $P(\theta)$ is the probability of the occurrence of shimmering-active neighbours $N_{r e l}$ at the respective neighbourhood angle $\theta$. The relative proportion of active agents, including both bucket bridgers and generators, as derived from the tri-state IAR model, is $N = (\langle k_B \rangle + \langle k_G \rangle)a(t)$. To get an intuition of the respective neighbourhood angle, previous work has studied this parameter from the waves travelling from the right to the left~\cite{speedingsocialwaves}.
\newline

 Fig.~\ref{fig1} gives a visual representation of this angle. The directed triggered hypothesis states that the information transfer amongst agents contributes to less than $5\%$ of wave propagation in the main direction of the wave~\cite{speedingsocialwaves,decisionMaking,StereoMotion,SocialRepel,BeeVibrate}. Using the wave strength function the directed triggered hypothesis can be proven by demonstrating the strength of the information transfer in various directions. In turn the probability of information transfer is equivalent to the transmissibility $\lambda$ defined in the proposed model which will be given as a function of the probability of shimmering-active agents~\cite{speedingsocialwaves,decisionMaking,StereoMotion}. \\  

The transmissibility $\lambda$ of the system, as derived from the directed triggered hypothesis, is a function of $P(\theta)$: 
 
 \begin{equation}\label{7}
 \lambda=\frac{\frac{P\left(\theta\right)}{r_{2}}}{1+\frac{P\left(\theta\right)}{r_{2}}}. 
 \end{equation}

The derivation for Eq.(\ref{7}) can be found in the \hyperlink{lambda}{\textbf{Appendix}} section. To prevent a non-zero wave strength to occur in the direction of the wave, a small angle tolerance of $\epsilon$ is added to the angle $\theta$. As such, the neighbourhood angle is now $\theta+\epsilon$. This substitution for transmissibility can be used in the wave strength function to show the strength of information transfer in a desired direction from a wave originating from any arbitrary direction. Although the right to left convention has been used in many studies~\cite{speedingsocialwaves,decisionMaking,StereoMotion}, by assigning the coordinate axis in a different direction such that the blue arrow in Fig.~\ref{fig1} points to the oncoming wave, by symmetry the transmissibility as a function of the respective neighbourhood angle can still be applied given any coordinate reference. 

\section{Computational Simulation of Bee Shimmering Behaviour}

\subsection{State Interaction Model}
The state interaction model Eq.(\ref{1}) is simulated to visualise the minor and major shimmering  phenomena which can be seen in the \hyperlink{video1}{\textbf{Supplementary Video}}.  To visualise both minor and major shimmering interactions using simulation experiments with the aid of the EoN Python package~\cite{Miller2019}, we initialise the total number of agents $N$ using a grid composed of $(x_{max}, y_{max}) = (\sqrt{N}, \sqrt{N})$, where $x_{max}$ and $y_{max}$ represent the maximum number of agents per $x$ and $y$ directions, respectively. We denote $A(0)$ and $I(0)$ as the initial number of active and inactive agents respectively at time $t=0$. The initial number of active agents $A(0)$ are defined as the generator agents as portrayed in Fig.(\ref{fig1}), these agents are initialised randomly on the grid during simulation. After initialising the corresponding rates $r_{1}$ through $r_{5}$ and designating the generator agents $A(0)$ to initiate the wave, the shimmering interactions transpire. At each time step $t$, the interactions described in Eq. (\ref{q}) occur based on the agents' states (active, inactive, or relapsed) in the subsequent time step $t+1$. The process is iterated and recorded concurrently, displaying the number of agents in each state within an animated graph. The simulation results of minor shimmering and major shimmering examples are shown in the \hyperlink{video1}{\textbf{Supplementary Video}}. In the state interaction framework, we define the neighbouring agents as to whom they are connected to. In the simulation, each agent is next to 4 other agents, with the exception of agents connected at the corners and on the edges of the visual display. We only consider agents connected up, down, left, and right of every corresponding agent and not diagonals. Although this parameter could be altered, it is currently kept unchanged throughout the whole simulation. Furthermore, spatial coordinates are irrelevant as interactions only occur depending on the neighbourhood of the agents' states. It is computationally found when $R_{0} \le 3.5$, the rate of removing the active agents is much greater than the rate of inducing active agents, therefore it implies minor shimmering. When $R_{0}>3.5$, the wave propagation achieves a greater distance, hence major shimmering.

\subsection{Tri-State IAR Model}
The tri-state IAR model is simulated by presenting approximate solutions to the differential equations using the Runge-Kutta method ODE45 function in Matlab~\cite{ODE}. The results obtained to validate the proposed theorem are various examples to demonstrate the behaviour when $r_{2}<\delta$, and when $1-e^{-\delta}<{r_{2}}<\delta$, respectively, and how these conditions relate to major shimmering, bucket bridging and saltatoric processes for minor shimmering. Each parameter can be altered, which may slightly change the graphs. However, the graphs are only representative examples of what would typically happen when the most impactful variables are changed to adapt to different shimmering scenarios. The results are further elaborated below. For the simulations of the tri-state IAR model, we denote $a(0)$ and $i(0)$ as the initial proportions of active and inactive agents respectively at $t=0$. 

\begin{figure}[htb]
\begin{center}
    \includegraphics[width=1\textwidth]{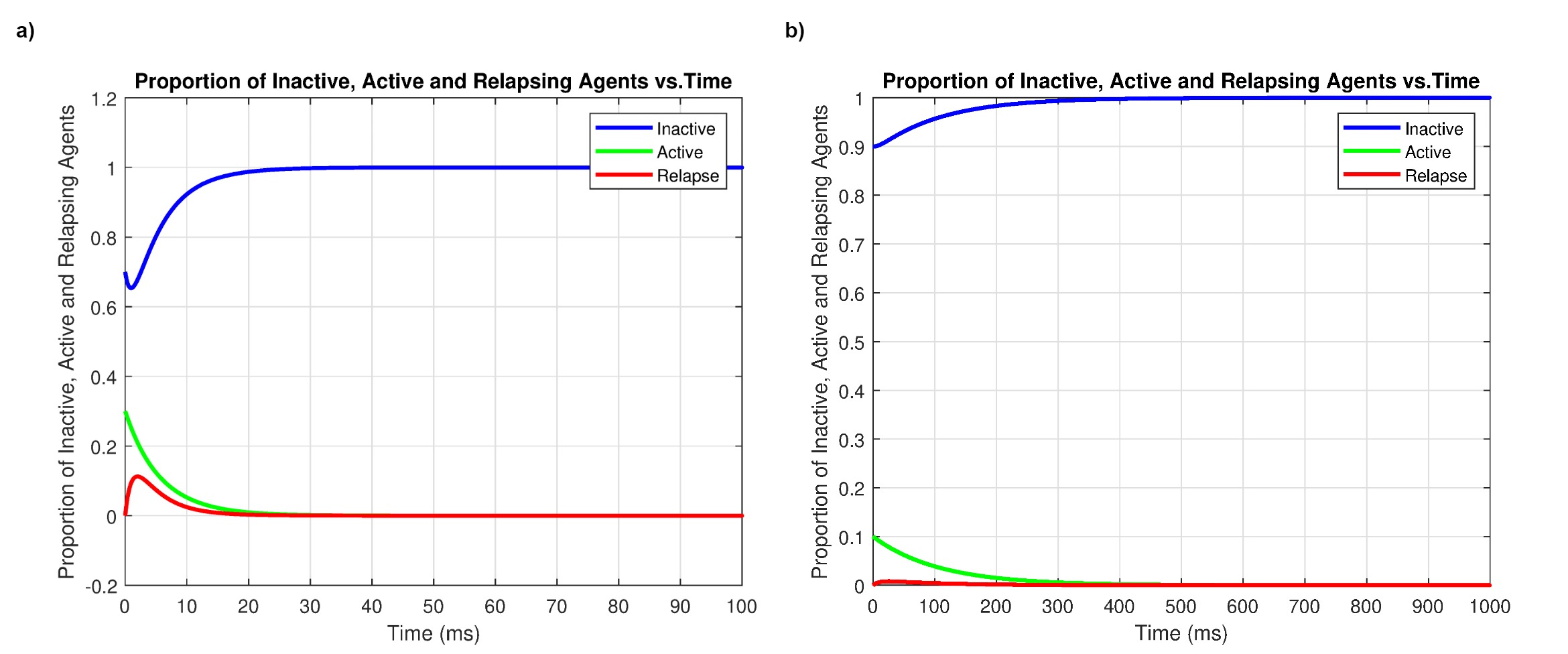}
\end{center}
  \linespread{1}\selectfont
\caption{{\bf Proportion of Inactive Active and Relapsing agents for Minor Shimmering}
Figure (a) displays a saltatoric process plot of the proportion of Inactive, Active and Relapsing agents over time by choosing the initial conditions $a(0)=0.3$ and $i(0)=1-a(0)$, $r_{2}=0.2$, $\lambda=0.3$, $r_{3}=0.9$, $\left\langle k_{B}\right\rangle=8$, $\left\langle k_{G}\right\rangle=0.1$, $\alpha=0.9$ and $\beta=0.9$. Figure (b) displays a bucket bridging plot of the proportion of Inactive, Active and Relapsing agents over time by choosing the initial conditions $a(0)=0.1$ and $i(0)=1-a(0)$, $r_{2}=0.01$, $\lambda=0.001$, $r_{3}=0.1$, $\left\langle k_{B}\right\rangle=8$, $\left\langle k_{G}\right\rangle=1$, $\alpha=0.9$ and $\beta=0.9$}
\label{fig3}
\end{figure}

\begin{figure}[htb]
\begin{center}
    \includegraphics[width=1\textwidth]{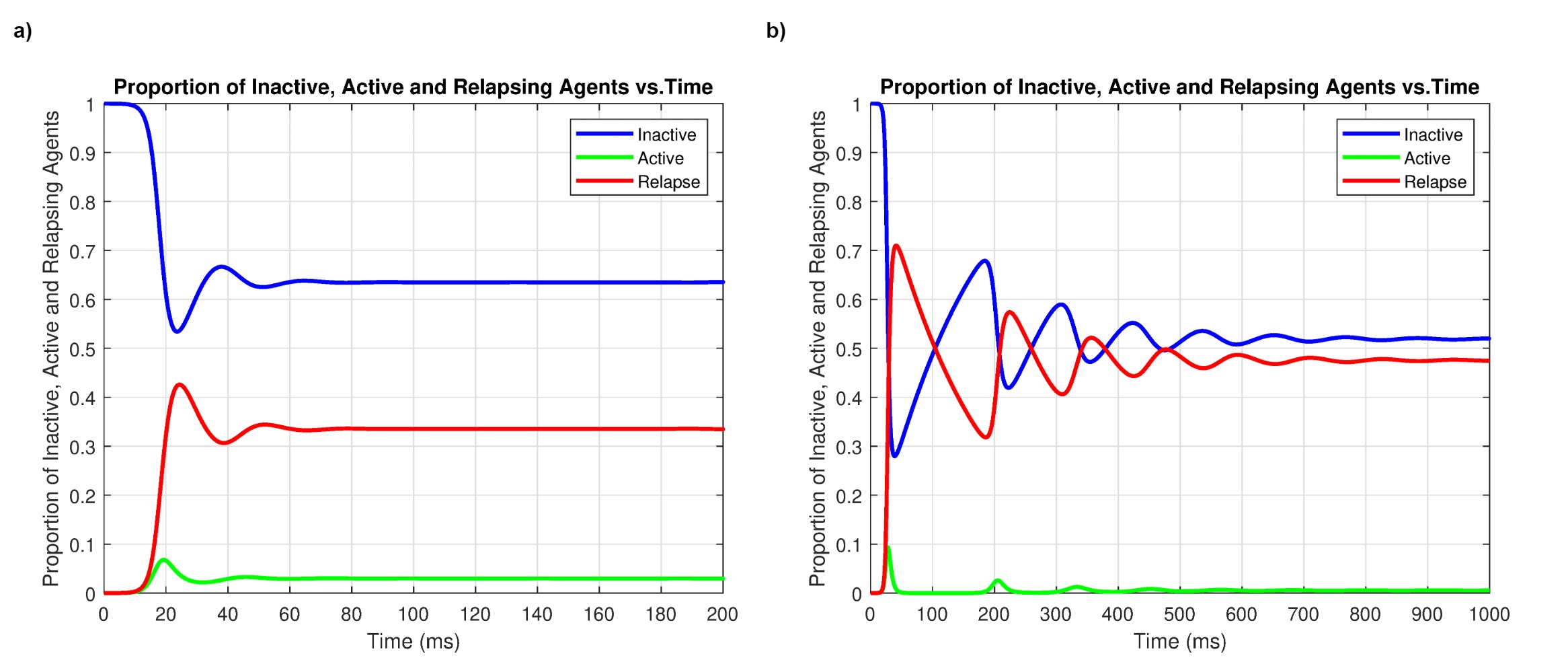}
\end{center}
  \linespread{1}\selectfont
\caption{{\bf Proportion of Inactive Active and Relapsing agents for Major Shimmering}
Part (a) displays the proportion of inactive, active, and relapsing agents over time, calculated using the following initial conditions: $a(0)=0.00001$, $i(0)=1-a(0)$, $r_{2}=0.9$, $\lambda=0.8$, $r_{3}=0.1$, $\left\langle k_{B}\right\rangle=10$, $\left\langle k_{G}\right\rangle=1$, $\alpha=0.9$, and $\beta=0.9$ without further subsequent waves. Part (b) shows the same information, but with initial conditions of $a(0)=0.000001$, $i(0)=1-a(0)$, $r_{2}=0.5$, $\lambda=0.8$, $r_{3}=0.006$, $\left\langle k_{B}\right\rangle=8$, $\left\langle k_{G}\right\rangle=1$, $\alpha=0.9$, and $\beta=0.6$, including daughter waves.}
\label{fig4}
\end{figure}

One of the initial assumptions in this study states that the pre-stroke phase is not considered. However, both saltatoric and bucket bridging processes can occur at this phase, and the minor shimmering processes are only analyzed in a limited time domain. According to Proposition 2, the system's steady states appear to converge to $(x_{1},x_{2})=(1,0)$ when $r_{2}>\delta$. This convergence takes place over a period of time, as shown in Fig.\ref{fig3} (a) and (b), where the results suggest minor shimmering due to $r_{2}>\delta$. The peaks of the wave can be seen in the first 0-100ms of the shimmering process. Minor shimmering is characterized by a relatively quick dispersion of the wave ~\cite{speedingsocialwaves}, which is facilitated by a low transmissibility value of $\lambda$. Fig.\ref{fig3} (a) shows quicker convergence to equilibrium compared to Fig.\ref{fig3} (b) because of the higher relapse rate $r_{2}=0.2$ in Fig.\ref{fig3} (a) compared to $r_{2}=0.01$ in Fig.\ref{fig3} (b), which speeds up the rate at which active agents reach the relapse state. The effect of a significantly higher $r_{3}=0.9$ in Fig.\ref{fig3} (a) compared to $r_{3}=0.1$ in Fig.\ref{fig3} (b) also contributes to the faster return of relapsing agents to the inactive state.  Although the transmissibility $\lambda$ is lower in Fig.\ref{fig3} (a) than in Fig.\ref{fig3} (b), the effect of lowering $r_{2}$ and $r_{3}$ still supersedes the effect of changing $\lambda$ as long as $r_{2}>\delta$ holds.  The initial conditions of $a(0)=0.3$ in Fig.\ref{fig3} and $a(0)=0.1$ in Fig.\ref{fig4} do not affect the condition of $r_{2}>\delta$. However, a higher $a(0)$ slightly increases the convergence time, but the effects of $r_{2}$ and $r_{3}$ still supersede the initial conditions. The quicker dissipation of the wave in Fig.\ref{fig3} (a) suggests a saltatoric process which is known to have a greater propagation wave velocity ~\cite{speedingsocialwaves}. The slower settling time in Fig.\ref{fig3} (b) implies a bucket bridging process of minor shimmering. It was initially assumed that there are only conditions for major and minor shimmering, but as shown in Fig.\ref{fig3} (a) and (b), changing the values of $r_{2}$ and $r_{3}$ can alter the settling time for equilibrium, leading to either saltatoric or bucket bridging processes for minor shimmering. Fig.~\ref{fig3} suggests minor shimmering due to $r_{2}>\delta$. Further examples of minor Shimmering results can be seen in the \hyperlink{supplementary}{\textbf{Supplementary Information}} with a varying initial number of active agents.
\newline

Given the condition $1-e^{-\delta}<{r_{2}}<\delta$ from \hyperlink{Prp3}{\textbf{Proposition 3}}, two examples of major shimmering types are shown in Fig.\ref{fig4}. These results suggest that with the given parameter choices, there is only one large shimmering wave under the condition specified. The shape of the shimmering wave is primarily determined by the increase in transmissibility $\lambda$, leading to ${r_{2}}<\delta$. A key difference between the results shown in Fig.\ref{fig3} and Fig.~\ref{fig4} is that the steady states can be anywhere in the interval $[0,1]\times(0,1]$, and they do not converge to $(x_{1},x_{2})=(1,0)$, because of the choice of ${r_{2}}<\delta$.

In Fig.\ref{fig4} (a), the inactive and relapsed states eventually converge to a higher value than the active state. This makes sense, as after the major wave, many agents remain in a relapsed state, and only a small proportion become active as the wave continues to spread at a steady rate. However, under the assumption of considering only the major wave, the equilibrium points do not last as long as $t\rightarrow\infty$ in reality. Other actions of minor or major shimmering may occur afterwards. Based on the assumption that bees could fly away given a longer period of time\cite{speedingsocialwaves,decisionMaking,StereoMotion,BeeVibrate,SocialRepel,BeeBook,BeeDefStrat,Assam,BeeSelfAssem,BeeKinematics}, after at least 1000 ms, it is assumed that these equilibrium points should not hold, resulting in a change in the total population of bees.
\newline

From \hyperlink{Prp3}{\textbf{Proposition 3}}, by decreasing $r_{2}=0.5$, $r_{3}=0.006$, $\left\langle k_{B}\right\rangle=8$ and $\beta=0.6$ from $r_{2}=0.9$, $r_{3}=0.2$, $\left\langle k_{B}\right\rangle=10$ and $\beta=0.9$ as shown in Fig.~\ref{fig4} (b) produce an oscillatory behaviour which eventually converges to an equilibrium in $[0,1]\times[0,1]$. The number of active agents at the peak decreases rapidly, however, smaller oscillations can be seen. This is still classified as major shimmering as the states do not converge to $(x_{1},x_{2})=(1,0)$ with $r_{2}<\delta$. However, the oscillations suggest that smaller waves known as daughter waves ~\cite{speedingsocialwaves} can occur. This phenomena usually happens when the predator does move away from the hive but is still within the vicinity. This type of shimmering can trigger further major shimmering if the predator returns closer to the hive. One of the parameters which aided such oscillations is a decrease of $r_{3}$ under the major shimmering conditions as it is a function of the number of inactive and relapsing agents. See the \hyperlink{supplementary}{\textbf{Supplementary Information}} for more examples of major shimmering results with varying initial number of active agents.
\newline

\subsection{Wave strength function}

\begin{figure}[htb]
\begin{center}
    \includegraphics[width=1\textwidth]{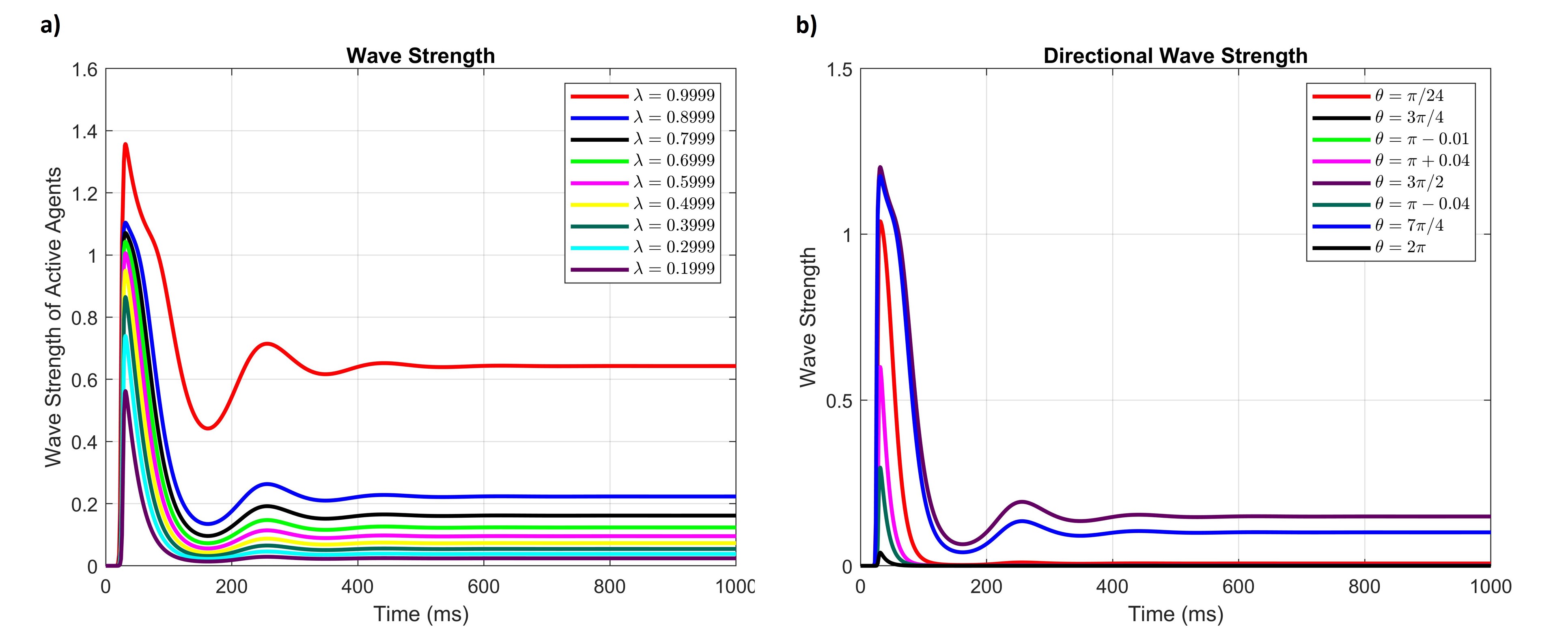}
\end{center}
  \linespread{1}\selectfont
\caption{{\bf Wave Strength Function results}
Plot (a) shows the wave strength function over time with varying $\lambda$ in intervals of 0.1 and choosing $a(0)=1\times 10^{-13}$, $i(0)=1-a(0)$, $r_{2}=0.05$, $r_{3}=0.001$, $\left\langle k_{B}\right\rangle=10$, $\left\langle k_{G}\right\rangle=4$, $\alpha=0.9$, $\beta=0.03$. Plot (b) shows the strength of the wave moving from right to left using the same parameters as (a) in a given direction $\theta$ with an angle tolerance of $\epsilon=0.01$ radians}. 
\label{fig5}
\end{figure}

Different wave strengths were generated using the wave strength function~\cite{speedingsocialwaves,decisionMaking,StereoMotion}. The results are shown in Fig.\ref{fig5}.

The results in Fig.\ref{fig5}(a) show wave strengths similar to those measured in empirical studies on bees\cite{speedingsocialwaves, decisionMaking, StereoMotion, SocialRepel, BeeVibrate}. These empirical studies measure wave strength using camera scalability of luminance changes, with relative units that vary between studies and remain non-dimensional~\cite{speedingsocialwaves, decisionMaking, StereoMotion}. The shape and characteristics of the graphs in these studies are consistent with the results of this study. The wave strength in this study is relative to the total rate of conversion from active to inactive agents. The luminance changes in previous studies~\cite{speedingsocialwaves, decisionMaking, StereoMotion} are equivalent to the total rate of conversion from inactive to active states. The parameters of the wave strength function can be altered to change the shape and magnitude of the major shimmering waves.
\newline

Fig.\ref{fig5} (a) shows the plot of the wave strength function derived from Eq.(\ref{5}) that takes into account only the post-stroke phase of shimmering. The wave strength model shares similarities with neuronal bursting and action potential models\cite{SelfOrgBook,ComplexityBook}. It exhibits equivalent points of stimulation, depolarisation, repolarisation, and a refractory period that converge to the resting state, similar to the Hindmarsh rose model~\cite{NeurBurst1,NeurBurst2,NeurBurst3,NeurBurst4,NeurBurst5,NeurBurst6,NeurBurst7,NeurBurst8,NeurBurst9}. In the case of bee shimmering, stimulation occurs when a wasp is at the closest proximity to the surface of the nest, triggering the bees to enter an active state, equivalent to depolarisation~\cite{NeurBurst1,NeurBurst3,NeurBurst4,NeurBurst5}. After reaching the peak, repolarisation occurs, leading to a refractory period, which can be seen as troughs in Fig.\ref{fig5} (a), where the bees need to recharge and settle into an inactive state. A few oscillatory periods may be seen when $\lambda$ is relatively high, indicating smaller daughter waves immediately following the major shimmering wave. The lower $\lambda$, the less likely follow-up daughter waves will occur. However, as the model only takes into account the major wave and smaller wavelets that occur immediately after, the resting state should not last for more than 1 second\cite{speedingsocialwaves,decisionMaking,StereoMotion}. Note that if greater time is taken into consideration, factors such as wildlife actions (e.g. bee movements), large animals, major shimmering waves, and other environmental scenarios would affect the next immediate behaviour of agents within the system.
\newline

One property of interest in this study is the direction of how powerful the wave strength is. It has been proven that less than $5\%$ of agents transfer the information in the same direction as the wave~\cite{speedingsocialwaves}.
\newline

\subsection{Wave strength function with directional control}

Consider the wave strength with respect to angles of $2\pi$ and $\pi$ in the coordinate frame assigned in Fig.~\ref{fig1}. Consider values which enable the major shimmering conditions shown in \hyperlink{Prp3}{\textbf{Proposition 3}} to hold true. It has been shown that within a tolerance of $\epsilon = 0.01$, the wave at peak will be less than $5\%$ in the direction of shimmering. By choosing a tolerance angle of $\epsilon = 0.01$, as shown in Fig.~\ref{fig5} (b), when angle $\theta$ is at $\pi$ when the waves move from right to left, the strength of the information transfer is very small, close to 0. This is opposed to when the angle $b_{N h}$ is close to $\pi/2$ or at $3\pi/2$, when maximum wave strength transfer can occur. This  further validates the fact that information transfer is strongest either directly upwards or downwards from a wave approaching the left from the right. A minimum information transfer for bees occurs when the angle $b_{N h}$ is closer to $\pi$ away from the active agent (directly left from a wave coming from the right). It also makes sense that there is near zero information transfer when the angle $\theta$ is at 0 or $\pi$, meaning the wave does not rebound back to the direction it came from. Overall, information transfer within bee shimmering reaches its full potential upwards and downwards if a wave moves from right to left. Eq.\ref{6} also implies that the probability for an active agent to converse an inactive agent is significantly much greater inducing agents who are upwards or downwards the relative agent. This information, when combined with the IAR model, can be used to predict the strength of shimmering information transfer across bee hives and detect different types of shimmering behaviours based on combinations of information transfer and steady states.

\section{Conclusions}

% 1 or 2 short paragraphs of discussion

This study presents the first analytical mathematical model of bee shimmering to classify behaviours ranging from saltatoric and bucket bridging processes to the major shimmering wave in the post-stroke phase within the first second. Initially, the state interaction model was derived and simulated in animation to visualise basic major and minor bee shimmering behaviours. Although both saltatoric and bucket bridging processes were initially considered as minor shimmering, the later assumptions based on redundant dynamics and the conditions imposed in the three propositions expanded the scope of shimmering behaviours considered. This expansion was due to the introduction of additional parameters such as the agent type and transmissibility $\lambda$, and their effects on rates $r_2$ and $r_3$. To test the validity of the assumptions for redundant dynamics, a wave strength function was introduced and compared with previous studies~\cite{speedingsocialwaves,decisionMaking,StereoMotion,SocialRepel,BeeVibrate}. In addition to these comparisons, the study supports the direct-trigger hypothesis stated in~\cite{speedingsocialwaves} that only 5$\%$ of the main major shimmering wave travels in the same direction. The wave strength function was derived using Eq.(\ref{6}) and the computational simulation results were plotted in Fig.\ref{fig5}(b). This gave further validation to the redundant dynamics assumptions used in the IAR model and stability analysis for major shimmering in \hyperlink{Prp3}{\textbf{Proposition 3}}. The IAR model provides a prediction of the wave strength that is consistent with literature\cite{speedingsocialwaves,decisionMaking,StereoMotion}. This study draws inspiration from multiple studies to model the interactions and changes of different bee shimmering states~\cite{speedingsocialwaves,decisionMaking,StereoMotion,BeeVibrate,SocialRepel,BeeBook,BeeDefStrat,Assam,BeeSelfAssem,BeeKinematics,MexicanWaves,MexicanWaves2,MexicanWaves3,AnimalInter,Starling1,SelfOrgBook,SyncBook,ComplexityBook,Starling2,Starling3,Starling4,Starling5,RecNN,Percolation1,Percolation2,Percolation3}. The IAR model has the ability to model the information transfer among surrounding agents and its features were determined by the propositions and proofs presented in this study. 

This study provides a significant first step towards analytically modelling the bee shimmering phenomena through the lens of non-linear dynamical systems. Although the novel IAR model yields promising results by incorporating a multitude of state interaction assumptions, future studies may further investigate additional aspects of bee shimmering through a more complex networks approach. One example is the pre-stroke phase and the various waves that follow after the major shimmering wave in different scenarios~\cite{speedingsocialwaves,decisionMaking,StereoMotion,BeeVibrate,SocialRepel,BeeBook,BeeDefStrat,Assam,BeeSelfAssem,BeeKinematics} which could be seen as phase transition problem. In addition, the effects of changes in the bee population over time and the speed of the shimmering waves may also be addressed as a node percolation problem. By considering the relevant speeds and their relationship to the number of agents and information transfer, it could be possible to accurately classify the types of waves analytically through network structure.

\section*{Appendix}

\subsection*{Redundant dynamics and further assumptions}

The state interaction model provides an initial foundation for the IAR model based on the initial assumptions. However, they do not account for the network structure of the three different types of agents yet. Further modifications are made from further assumptions which can then be made to consider the average number of agent type around a particular agent (average degrees). Therefore, a mean-field analysis can be taken to redefine the dynamics of the chain tail agent.

In Fig.~\ref{fig1} there is a depiction of how information is transferred amongst the three types of bee agents. To take account of the types of agents, we define new variables describing the average number of bucket bridger and generator neighbours a randomly selected agent has. This is known as the average degree. The average degrees of the bucket bridger and generator agents would be introduced as, $\left\langle k_{B}\right\rangle$ and $\left\langle k_{G}\right\rangle$ respectively. In turn $\left\langle k_{C}\right\rangle$ would be the average degree of the chain tail agents. However from the definitions of what chain tail agents are capable of \cite{speedingsocialwaves,decisionMaking,StereoMotion}, and illustrated in both Fig.~\ref{fig1} and Fig.~\ref{fig2}, these agents are redundant. In other words, they do not do anything to pass the propagation of information, they are termed as impaired agents, as they prevent information from travelling. Thus the variables $\left\langle k_{B}\right\rangle$ and $\left\langle k_{G}\right\rangle$ should be introduced, whereas the chain tail agents could be modelled as a percolation \cite{Percolation1,Percolation2,Percolation3} of damaged nodes in the network. Therefore the network of agents within the bee shimmering time frame only consists of bucket bridgers and generator agents, where the failed bucket bridger and generator agents are chain tail agents. When $r_{1}$ and $r_{4}$ fail, these agents are not induced into the active state and ultimately stay in the inactive state.

Currently, the distribution of these types of agents is random \cite{speedingsocialwaves,decisionMaking,StereoMotion}. This means that there is an assumption that by selecting an agent randomly there is a chance that it could be connected to bucket bridgers as well as a chance connected to a generator agent. We assume that the bee hive is homogeneously densely packed for bees and that generator agents could be initialised anywhere and randomly in the simulation. Therefore the dynamics of interactions shown in  Fig.~\ref{fig2} (a) (b) (c) and (d) can be simplified with a replacement of chain tail agents. 
\newline

\hypertarget{IARmodel}{\subsection*{Tri State IAR model derivation}}

By substituting the equations for $r_{1}$ and $r_{4}$ into the state interaction model, we obtain the tri state IAR model Eq.(\ref{2}), and Eq.(\ref{3}). Upon deriving this model, it is now of interest to determine its steady states, invariance properties and stability conditions as well as the physical implications.

\subsection*{Invariance and stability analysis}

By re-naming ${x}_{1}=i$, ${x}_{2}=a$, and by letting $\gamma=\alpha\left\langle k_{B}\right\rangle\ln(1-\lambda)^{-1}$ and $\delta=\beta\left\langle k_{G}\right\rangle\ln(1-\lambda)^{-1}$, Eq.(\ref{11})-Eq.(\ref{12}) can be re-written as:

\begin{equation}\label{16}
\left(
\begin{array}{c}
  \dot{x}_1 \\ \dot{x}_2
\end{array}
\right)=
\left(
\begin{array}{c}
   -x_{1} x_{2}\left(1-e^{- \gamma x_{2}}\right)+r_{3}\left(1-x_{1}-x_{2}\right)- 
x_{1}\left(1-e^{-\delta x_{2}}\right) \\
-r_{2} x_{2}+x_{1}\left(1-e^{-\delta x_{2}}\right)
\end{array}
\right)=:f(x_1,x_2)
\end{equation}

Let $\mathcal{S}=[0,1]\times [0,1]$ be the unit square. Note that it follows from the previous subsection that we need $0\le x_1\le 1,~0\le x_2\le 1$, which means that $\mathcal{S}$ is a positively invariant set \cite{DynamicalSystems1} for Eq.(\ref{16}). The following proposition provides conditions for invariance.

 \hypertarget{proof1}{\textbf{Proof for Proposition 1.}} 

\begin{proof}
  We consider the boundary of $\mathcal{S}$ and ensure that the vector field $f$ defined by Eq.(\ref{16}) is not pointing out of $\mathcal{S}$.
  
  First, when $x_1=0$, we have $\dot{x}_1=r_3(1-x_2)\ge 0$ for $0\le x_2\le 1$, which is as required.
  
  Second, when $x_1=1$, we have $\dot{x}_1=-x_2(1-\mathrm{e}^{-\gamma x_2})-r_3x_2-(1-\mathrm{e}^{-\delta x_2})\le 0$, as each of the individual terms are non-positive for $0\le x_2\le 1$.
  
  Third, when $x_2=0$, we have $\dot{x}_2=0$. 
  Finally, when $x_2=1$, we have $\dot{x}_2=-r_2+x_1(1-\mathrm{e}^{-\delta})$, which is non-positive for all $0\le x_1\le 1$ if and only if $r_2\ge 1-\mathrm{e}^{-\delta}$. This establishes our claim.
\end{proof}

Note that we have $f(1,0)=(0,0)$, which means that the minor shimmering point $(1,0)$ is always a steady state for (\ref{16}). Moreover, it is straightforwardly shown that the linearisation of (\ref{16}) around $(1,0)$ is given by 
\begin{equation}
  \label{17}
    \dot{\eta}=
    \left(
      \begin{array}{cc}
        -r_3&-\delta-r_3 \\ 0&\delta-r_2
      \end{array}
    \right)\eta
\end{equation}

\hypertarget{proof2}{\textbf{Proof for Proposition 2.}} 

\begin{proof}
  When $r_2>\delta$, it is clear that all eigenvalues of the matrix in (\ref{17}) are negative, which means that $(1,0)$ is a locally stable steady state for (\ref{4}).To show global asymptotic stability on $\mathcal{S}$, first note that when $x_2=0$, we have $\dot{x}_2=0$ and $\dot{x}_1=r_3(1-x_1)>0$ for $0\le x_1<1$, which means by LaSalle's Invariance Principle \cite{DynamicalSystems2} that $x_1(t)\to 1$ as $t\to+\infty$. Secondly, if $0<x_2\le 1$, we have
  \[
    \dot{x}_2=-r_2x_2+x_1(1-\mathrm{e}^{-\delta x_2})<-r_2x_2+1-\mathrm{e}^{-\delta x_2}<0
  \]
  as $-r_2x_2+1-\mathrm{e}^{-\delta x_2}=0$ when $x_2=0$ and $\frac{\partial}{\partial x_2}\left(-r_2x_2+1-\mathrm{e}^{-\delta x_2}\right)<0$ for $0<x_2\le 1$. Once again by LaSalle's Invariance Principle \cite{DynamicalSystems2}, this means that solutions converge to largest invariant subset that is contained in $\dot{x}_2=0$, which can be easily shown to be the minor shimmering point $(1,0)$

\end{proof}

It may be shown that the dynamical system (\ref{16}) undergoes a transcritical bifurcation at the minor shimmering point $(1,0)$ when $r_2=\delta$ \cite{DynamicalSystems1}. This means that when $r_2$ is reduced below $\delta$, the minor shimmering point becomes unstable (though it is a saddle point, so solutions starting on its stable manifold $x_2=0$ will still converge to the minor shimmering point), and a stable major shimmering point is "born". It may be shown that for values of $r_2<\delta$ close to $\delta$ this major shimmering point is located on the interior of $\mathcal{S}$  and that it is locally asymptotically stable. The following proposition actually shows that the major shimmering point is unique and globally asymptotically stable on ${\mathcal{S}}^*=[0,1]\times(0,1]$ for all $1-\mathrm{e}^{-\delta}\le r_2<\delta$.

\hypertarget{proof3}{\textbf{Proof for Proposition 3.}} 

\begin{proof}
  Setting the second component of $f$ in Eq.(\ref{4}) equal to zero and solving for $x_1$, gives that at steady states of Eq. (\ref{4}) we have that 
  \begin{equation}
      \label{18}
      x_1=g(x_2)=\frac{r_2x_2}{1-\mathrm{e}^{-\delta x_2}}
  \end{equation}
  
  Note that by L'H\^{o}pital's Rule we have that $g(0)=\frac{r_2}{\delta}<1$, while also $g(1)=\frac{r_2}{1-\mathrm{e}^{-\delta}}>1$.Moreover, as the numerator of $g(x_2)$ is a strictly increasing function of $x_2$, the denominator is a strictly decreasing function of $x_2$, and both numerator and denominator are strictly positive on $\mathcal{S}^*$, we have that $g(x_2)$ is a strictly increasing function of $x_2$.
  
  Next, setting the first component of $f$ in Eq. (\ref{16}) equal to zero and solving for $x_1$, gives that at steady states of Eq. (\ref{16}) we have that 
  \begin{equation}
      \label{19}
      x_1=h(x_2)=\frac{r_3(1-x_2)}{r_3+(1-\mathrm{e}^{-\delta x_2})+x_2(1-\mathrm{e}^{-\gamma x_2})}
  \end{equation}
  We then have that $h(0)=1$, $h(1)=0$ and using a similar argument as a above it may be shown that $h(x_2)$ is a strictly decreasing function of $x_2$.
  
  From the above we conclude that $k(x_2)=g(x_2)-h(x_2)$ is a strictly increasing function with $k(0)<0$ and $k(1)>0$, which means that there exists a unique $x_2^*\in (0,1)$ such that $k(x_2^*)=0$ and therefore on $\mathcal{S}^*$ there exists a unique steady state of Eq. (\ref{16}) at $(g(x_2^*),x_2^*)$.
  
  Having established the existence of a unique steady state on $\mathcal{S}^*$, we next show that this steady state is globally asymptotically stable on $\mathcal{S}^*$. By Poincar\'{e}-Bendixson theory \cite{DynamicalSystems2}, instability of the steady state would imply the existence of a globally asymptotically stable periodic solution on $\mathcal{S}^*$. Therefore, we can prove global asymptotic stability of the steady state on $\mathcal{S}^*$ if we can rule out the existence of periodic solutions on $\mathcal{S}^*$. The mechanism by which the unique steady state becomes unstable and a stable periodic solution appears is through a Hopf bifurcation \cite{DynamicalSystems1}. This means that the linearisation of (\ref{16}) at the steady state has two imaginary eigenvalues, meaning that the sum of the diagonal elements of the linearisation matrix is equal to zero. Therefore, we need the following equality to hold at a steady state:
  \begin{equation}
      \label{20}
      0=\frac{\partial f_1}{\partial x_1}+\frac{\partial f_2}{\partial x_2}=
      \mathrm{e}^{-\delta x_2}-r_3-r_2+x_2(\mathrm{e}^{-\gamma x_2}-1)+\delta x_1\mathrm{e}^{-\delta x_2}-1
  \end{equation}
  Solving (\ref{20}) for $x_1$ gives
  \begin{equation}
      \label{21}
      x_1=\frac{\mathrm{e}^{\delta x_2}}{\delta}\left(r_2+r_3+1-\mathrm{e}^{-\delta x_2}+x_2(1-\mathrm{e}^{-\gamma x_2})\right)>\frac{r_2\mathrm{e}^{\delta x_2}}{\delta}\ge
      \frac{r_2x_2}{1-\mathrm{e}^{-\delta x_2}}
  \end{equation}
  where the last equality follows from the fact that for $x_2\ge 0$ we have
  \[
    \mathrm{e}^{\delta x_2}-1-\delta x_2=\sum\limits_{k=0}^{\infty}\frac{1}{k!}(\delta x_2)^k-1-\delta x_2=\sum\limits_{k=2}^{\infty}\frac{1}{k!}(\delta x_2)^k\ge 0
  \]
  Now note that at the steady state we also need that Eq. (\ref{18}) holds, which clearly contradicts Eq. (\ref{21}). Therefore, we have ruled out the occurrence of a Hopf bifurcation, which means that the unique steady state on $\mathcal{S}^*$ is globally asymptotically stable on $\mathcal{S}^*$ for all $1-\mathrm{e}^{-\delta}\le r_2<\delta$
\end{proof}

\subsection*{Wave strength function with directional control}

The directional propagation of the wave describes the transfer of information depending on the agent participation (number of active agents) where the waves come from one particular direction to the other side. Following the directed triggered hypothesis \cite{speedingsocialwaves}, it is given that:

\begin{equation}\label{22}
P(\theta)=N_{r e l} \times \sin ^{2}\left(\theta\right)
\end{equation}

 $P(\theta)$ is the probability of the occurrence of shimmering-active neighbours $N_{r e l}$ at the respective neighbourhood angle $\theta$. As shown in the derivation for the tri-state IAR model, the relative proportion of active agents including both bucket bridgers and generators is derived as $N_{r e l}=(\left\langle k_{B}\right\rangle + \left\langle k_{G}\right\rangle) a(t)$. To get an intuition of the respective neighbourhood angle, previous work have studied this parameter from the waves travelling from the right to the left \cite{speedingsocialwaves}.
\newline

 Fig.~\ref{fig1} gives a visual representation of this angle. The directed triggered hypothesis states that the information transfer amongst agents contributes to less than $5\%$ of wave propagation in the main direction of the wave  \cite{speedingsocialwaves,decisionMaking,StereoMotion,Assam,SocialRepel}. Using the wave strength function the directed triggered hypothesis can be proven demonstrating the strength of the information transfer in various directions. In turn, the probability of information transfer is equivalent to the transmissibility $\lambda$ defined in the proposed model which will be given as a function of the probability of shimmering-active agents \cite{speedingsocialwaves,decisionMaking,StereoMotion}. Using the directed triggered hypothesis we proceed to derive the transmissibility $\lambda$ as a function of $P(\theta)$. \newline  \\  

\hypertarget{lambda}{\subsection*{Derivation for the transmissibility $\lambda$}}

%\label{secA1}

 Derivation for the transmissibility $\lambda$ of the system derived as a function of $P(\theta)$:
\newline

Consider $n$ intervals $\delta\tau$ of time $\tau$ where $\delta\tau<<1$ such that $n\delta\tau=\tau$. In reference to Fig.~\ref{fig1}, and during the interval of time $\delta\tau$, $1-r_{2}\delta\tau$ is the hypothetical probability that the active nodes do not enter the relapsed state in a single interval of $\delta\tau$. Considering the probability that none of these intervals are removed, the probability that the active agents do not enter the relapsed state during the time $\tau$ is:

\begin{equation}\label{23}
\lim _{\delta \tau \rightarrow 0}\left(1-r_{2} \delta \tau\right)^{\tau / \delta\tau}=
e^{- r_{2}\tau}
\end{equation}

 Thus the probability that an active node becomes relapsed at some arbitrary time $\tau^{\prime}<\tau$ is:

\begin{equation}\label{24}
P\left(\tau^{\prime}<\tau)=1-e^{-r_{2} \tau}\right.
\end{equation}

The probability that the active nodes turns into the relapsed state at time $\tau$ is:

\begin{equation}\label{25}
P(\tau)=\frac{d P\left(\tau^{\prime}<\tau)\right.}{d \tau}=\frac{d}{d \tau}\left(1-e^{-r_{2} \tau}\right)=r_{2}e^{-r_{2} \tau}
\end{equation}

Probability of the occurrence of inactive neighbours at time $\delta\tau<<1$ is:
\begin{equation}\label{26}
1-P(\theta) \delta \tau
\end{equation}

The probability that the active states do not converse with other agents in the interval $\delta\tau<<1$:

\begin{equation}\label{27}
\lim _{\delta \tau \rightarrow 0}\left(1-P(\theta) \delta \tau\right)^{\frac{\tau}{\delta \tau}} = e^{ -P(\theta)\tau}
\end{equation}

Thus the probability that the active states spread over an interval $\tau$ is:

\begin{equation}\label{28}
\lambda_{\tau}=1-e^{ -P(\theta)\tau}
\end{equation}

Given that the agents do not stay active for long, integrating over the distribution of time yields a relationship between the transmissibility of the system and the respective neighbourhood angle:

\begin{equation}\label{29}
\begin{array}{c}
\lambda=\int_{0}^{\infty} d \tau P(\tau) \lambda_{\tau} \\ 

\\
=\int_{0}^{\infty} d \tau\left(r_{2} e^{-r_{2} \tau}-r_{2} e^{\left(-r_{2}-P(\theta)\right) \tau}\right) \\

\\
\lambda=\frac{\frac{P(\theta)}{r_{2}}}{1+\frac{P(\theta)}{r_{2}}}  \\
\end{array}
\end{equation}

\bibliographystyle{unsrt}  
\bibliography{references}  %%% Remove comment to use the external .bib file (using bibtex).


% George Kour and Raid Saabne.
% \newblock Real-time segmentation of on-line handwritten arabic script.
% \newblock In {\em Frontiers in Handwriting Recognition (ICFHR), 2014 14th
%   International Conference on}, pages 417--422. IEEE, 2014.

% George Kour and Raid Saabne.
% \newblock Fast classification of handwritten on-line arabic characters.
% \newblock In {\em Soft Computing and Pattern Recognition (SoCPaR), 2014 6th
%   International Conference of}, pages 312--318. IEEE, 2014.

% \bibitem{hadash2018estimate}
% Guy Hadash, Einat Kermany, Boaz Carmeli, Ofer Lavi, George Kour, and Alon
% \newblock Estimate and replace: A novel approach to integrating deep neural
%   networks with existing applications.
% \newblock {\em arXiv preprint arXiv:1804.09028}, 2018.

% \end{thebibliography}

\section*{\hypertarget{supplementary}{Supplementary Information}}

\subsection*{\hypertarget{video1}{Supplementary Video: Bee Shimmering Simulation}} \url{https://youtu.be/eXbyVY3FJHY}

\noindent \textbf{Legend:} An example of minor shimmering waves is produced when we set a total population of $N=10,000$, $A(0)=3$, $I(0)=N-A(0)$, $r_1=4000$, $r_2=2300$, $r_3=3000$, $r_4=3000$, $r_5=0$, $R_0=3.043$. The horizontal axis shows the time in seconds which corresponds to the number of agent types on the vertical axis. As $R_0 \leq 3.5$ we see the minor shimmering behaviour occurring. The major shimmering waves are produced when we set the following values: $N=10,000$, $A(0)=3$, $I(0)=N-A(0)$, $r_1=4000$, $r_2=800$, $r_3=3000$, $r_4=3500$, $r_5=0$, $R_0=9.375$. The horizontal axis shows the time in seconds which corresponds to the number of agent types on the vertical axis. As $R_0>3.5$ we see the major shimmering behaviour occurring.
\newline

\begin{figure*}[htb]
\begin{center}
\hspace*{-1cm} 
    \includegraphics[width=1 \textwidth]{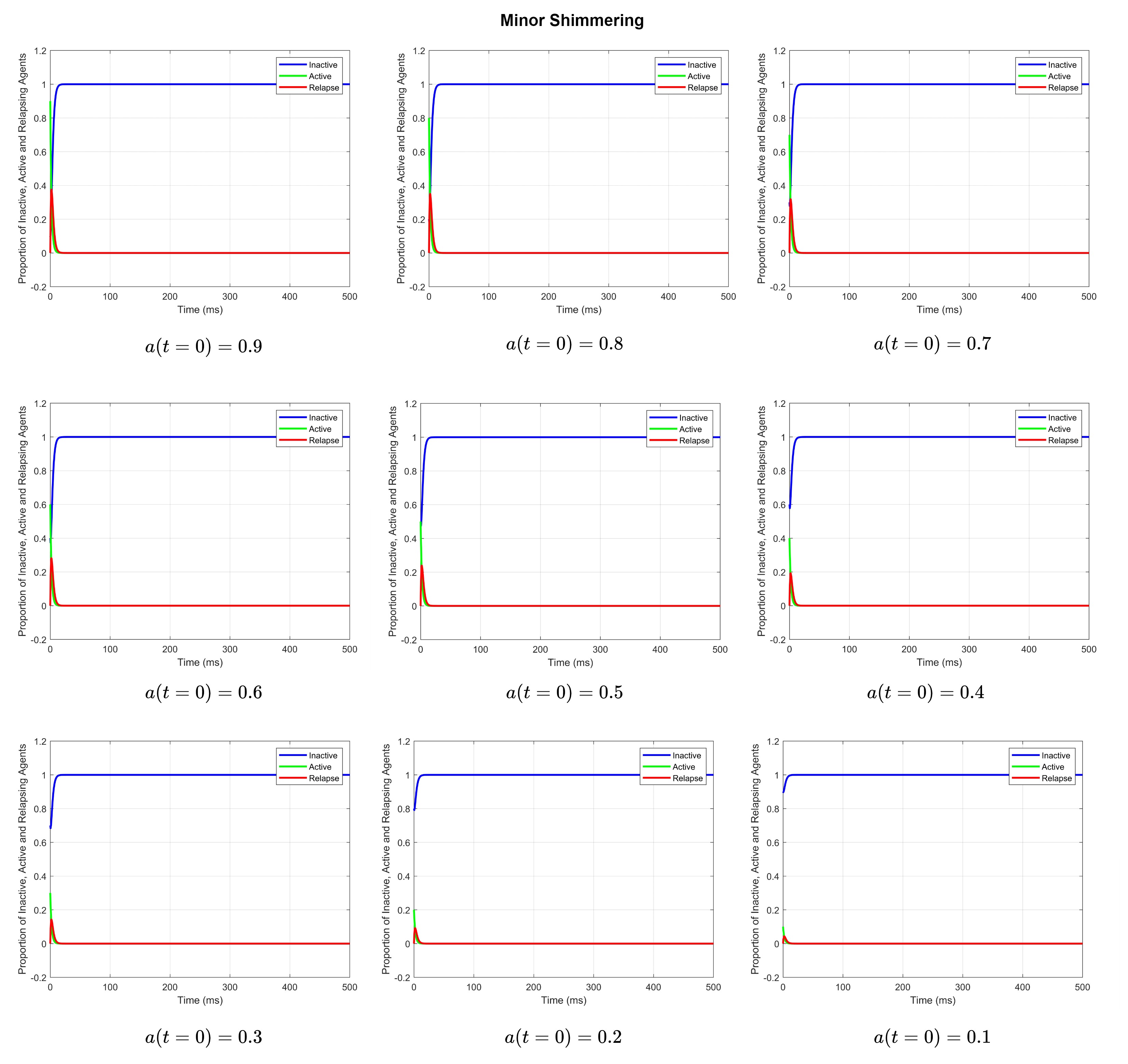}
\end{center}
  \linespread{1}\selectfont
\caption{\textbf{Supplementary Figure 1. Minor Shimmering Conditions} under a varying $a(0)$. $i(0)=1-a(0)$, $r_{2}=0.01$, $\lambda=0.001$, $r_{3}=0.1$, $\left\langle k_{B}\right\rangle=8$, $\left\langle k_{G}\right\rangle=1$, $\alpha=0.9$, $\beta=0.9$. By varying $a(0)$ we can see that the convergence still appears to satisfy $(i,a)=(1,0)$, under the condition that $r_{2}>\delta$.}
\end{figure*}

\begin{figure}[htb]
\begin{center}
\hspace*{-1cm} 
    \includegraphics[width=1\textwidth]{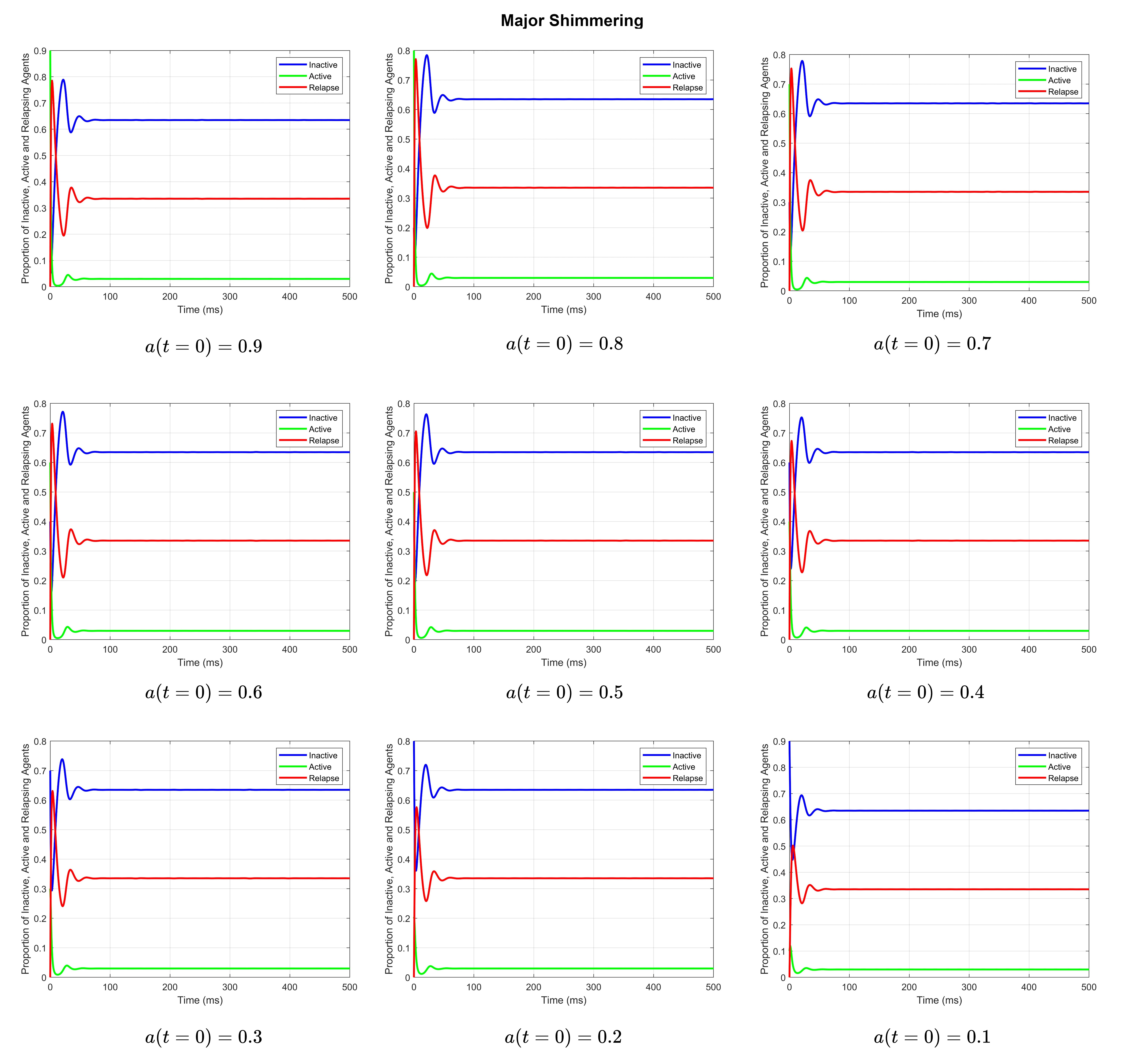}
\end{center}
  \linespread{1}\selectfont
\caption{\textbf{Supplementary Figure 2. Major Shimmering Conditions} under a varying $a(0)$. Where $i(0)=1-a(0)$, $r_{2}=0.9$, $\lambda=0.8$, $r_{3}=0.1$, $\left\langle k_{B}\right\rangle=10$, $\left\langle k_{G}\right\rangle=1$, $\alpha=0.9$, $\beta=0.9$. By varying $a(0)$, we can see that $i$ and $a$ satisfies a convergence inside of $\textbf{S}$. There is only one equilibrium point as $1-e^{-\delta}<{r_{2}}<\delta$ is set for major shimmering.}
\end{figure}

\subsection*{Minor Shimmering Examples}

By altering the initial proportion of active agents, further minor shimmering examples are simulated. The simulation results (Supplementary Fig.~1) demonstrated that the convergence time for the proportion of active and inactive agents settles down between $10-20ms$ for each graph, regardless of the variation in the initial proportion of active agents. This outcome further satisfies the condition wherein $r_{2}>\delta$ is required for minor shimmering to occur. A higher $a(0)$ marginally increases the convergence time, altering $a(0)$ does not result in the formation of major shimmering waves within this simulation.

\subsection*{Major Shimmering Examples}

Additional major shimmering examples (Supplementary Fig.~2) are observed below when changing the initial proportion of active agents. Similarly, in the case of major shimmering, the convergence time for the proportions of active and inactive agents settles between $50-75ms$ for each graph, regardless of variations in the initial proportion of active agents. The convergence time displays minimal differences from $a(0)=0.9$ to $a(0)=0.1$, adhering to the major shimmering condition of $1-e^{-\delta}<{r_{2}}<\delta$. The most significant variation lies in the magnitude of the proportion of inactive agents. With $a(0)=0.9$, the peak proportion of inactive agents is just below $0.8$, while it is slightly under $0.7$ when $a(0)=0.1$. A smaller proportion of inactive agents indicates a larger proportion of active and relapse agents at that point in time, implying a more extended major shimmering wave before any minor ones occur. In reality, shimmering at any given moment does not commence with a large proportion of active agents at $t=0$ for shimmering waves to be observed. These simulations demonstrate the implications of wave types that would otherwise transpire, which are not possible to capture in real bee shimmering.

\section*{Acknowledgements}

The authors gratefully acknowledge the helpful discussions and feedback given from A. Birn-Jeffery, Dr A. Buchan and the ARQ Robotics department at Queen Mary, University of London. A former undergraduate student J. Zahar aided in the development of the visual animation presented for the state interaction model.

\section*{Author Contributions Statement}

 N. Patel and K. Zhang conceived the study, developed the modelling and analytical framework, designed and performed the experimental simulations, and wrote and edited the manuscript. H. Huijberts developed the proof for all three propositions and aided in providing constructive feedback and support for the manuscript. All authors gave final approval for publication and agree to be held accountable for the work performed herein.

\section*{Data Availability}

The datasets used and analysed during the current study are available from the corresponding author on reasonable request.

\section*{Competing Interests}

The authors declare that they have no competing financial interests.

\end{document}